\documentclass[aps,onecolumn,prd,floatfix,preprintnumbers,superscriptaddress,notitlepage,nofootinbib,11pt]{revtex4-2}

\usepackage{amssymb,amsfonts,amsmath,mathtools,bm,xcolor,graphicx}
\counterwithout{table}{section}
\usepackage{hyperref}
\usepackage{array}
\hypersetup{colorlinks=true,linkcolor=blue,citecolor=blue,menucolor=black,urlcolor=blue,filecolor=blue}
\usepackage{orcidlink}
\usepackage[normalem]{ulem}

\graphicspath{{fig/}}
\renewcommand\arraystretch{1.25}
\allowdisplaybreaks

\newcommand{\eps}{{\varepsilon}}
 
\newcommand{\itp}{\affiliation{CAS Key Laboratory of Theoretical Physics, Institute of Theoretical Physics,\\
Chinese Academy of Sciences, Beijing 100190, China}}

\newcommand{\ucas}{\affiliation{School of Physical Sciences, University of Chinese Academy of Sciences, Beijing 100049, China}}

\newcommand{\peng}{\affiliation{Peng Huanwu Collaborative Center for Research and Education, International Institute for Interdisciplinary and Frontiers, Beihang University, Beijing 100191, China}}

\newcommand{\tud}{\affiliation{Institut für Kernphysik, Technische Universität Darmstadt, 
	64289 Darmstadt, Germany}}

\newcommand{\gsi}{\affiliation{ExtreMe Matter Institute EMMI and Helmholtz Forschungsakademie Hessen f\"ur FAIR (HFHF), GSI Helmholtzzentrum 
	für Schwerionenforschung GmbH, 64291 Darmstadt, Germany}}

\newcommand{\hiskp}{
\affiliation{Helmholtz--Institut f\"ur Strahlen- und Kernphysik (Theorie)\\ 
	and Bethe Center for Theoretical Physics, Universit\"at Bonn, D-53115 Bonn, Germany}}

\newcommand{\juelich}{\affiliation{Institute for Advanced
	Simulation (IAS-4), 
	Forschungszentrum J\"ulich, D-52425  J\"ulich, Germany}}

\newcommand{\tsu}{\affiliation{Tbilisi State University, 0186 Tbilisi, Georgia}}

\begin{document}

\title{Exploring Efimov  states in $D^*D^*D^*$ and $DD^*D^*$ three-body systems}

\author{Hai-Long Fu\orcidlink{0000-0002-1722-4145}}\email{fuhailong@itp.ac.cn}
\itp\ucas
\author{Yong-Hui Lin\orcidlink{0000-0001-8800-9437}}\email{yonghui.lin@tu-darmstadt.de}
\tud
\author{Feng-Kun Guo\orcidlink{0000-0002-2919-2064}}\email{fkguo@itp.ac.cn}
\itp\ucas\peng
\author{Hans-Werner~Hammer\orcidlink{0000-0002-2318-0644}}\email{Hans-Werner.Hammer@physik.tu-darmstadt.de}
\tud\gsi
\author{Ulf-G.~Mei{\ss}ner\orcidlink{0000-0003-1254-442X}}\email{meissner@hiskp.uni-bonn.de}
\hiskp\juelich\peng
\author{Akaki Rusetsky\orcidlink{0000-0002-4288-8672}}\email{rusetsky@hiskp.uni-bonn.de}
\hiskp\tsu
\author{Xu Zhang\orcidlink{0000-0002-3687-248X}}\email{zhangxu@itp.ac.cn}
\itp 

\begin{abstract}
The Efimov effect is an intriguing three-body quantum phenomenon. Searching for Efimov states within the realms of nuclear and hadronic physics presents a challenge due to the inherent inability of natural physical systems to exhibit adjustable two-body scattering lengths. In this study, we examine the potential existence of Efimov states in the $D^*D^*D^*$ and $DD^*D^*$ three-hadron systems. Utilizing a pionless effective field theory framework, we determine that the presence of Efimov states in the spectrum of the $D^*D^*D^*$ system is contingent upon the existence of an $(I,J)=(1,2)$ $D^*D^*$ two-body bound system. 
If only the $T_{cc}^+$ and its heavy quark spin partner $T_{cc}^{*+}$ exist while there is no near-threshold pole in all other $S$-wave $D^{(*)}D^*$ scattering amplitudes, no Efimov effect is expected in the $D^{(*)}D^*D^*$ systems.

\end{abstract}

\maketitle

\newpage

\section{Introduction}

Over the past two decades, a large amount of exotic hadronic states has been extensively investigated both experimentally and theoretically. These states demonstrate characteristics that cannot be adequately accounted for by the conventional quark model and investigations may lead to a deeper understanding of nonperturbative quantum chromodynamics (QCD), as discussed in numerous comprehensive reviews~\cite{Barabanov:2020jvn,Chen:2022asf,Brambilla:2019esw,Dong:2021bvy,Esposito:2016noz,Guo:2017jvc,Guo:2019twa,Hosaka:2016pey,Lebed:2016hpi,Liu:2019zoy,Mai:2022eur,Olsen:2017bmm,Yang:2020atz,Chen:2024eaq,Liu:2024uxn,Wang:2025sic}. 
The observation of complex structures in proximity to the thresholds of specific hadron pairs suggests that hadronic molecules provide a compelling interpretative framework for these phenomena~\cite{Guo:2017jvc}.
\footnote{This statement holds as long as there is no Castillejo-Dalitz-Dyson (CDD) zeros~\cite{Castillejo:1955ed} in the scattering amplitude in the vicinity of the threshold (see, e.g., Refs.~\cite{Kang:2016jxw, Dai:2023kwv,Song:2023pdq}). Until experimental evidence of such additional zeros is obtained, it is most natural to employ the Occam's razor principle to assume the smallest possible number of poles required by data and theory constraints in the scattering amplitude, as discussed in Ref.~\cite{Baru:2021ldu}.}

The concept of two-body hadronic molecules naturally extends to three-body systems (as recently reviewed in Ref.~\cite{Liu:2024uxn}), analogous to multi-nucleon atomic nuclei. The theoretical methodologies for three-body systems, particularly for multi-nucleon systems, have been developed since many years ago, as documented in Refs.~\cite{Blankenbecler:1965gx,Aaron:1968aoz,Blankenbecler:1961zz,Cook:1962zz,Fleming:1964zz,Grisaru:1966uev,Aaron:1973ca,Amado:1974za}. 
In recent years, various effective field theory (EFT) frameworks for
three-body systems in both finite and infinite volumes have been
proposed~\cite{Mai:2017vot,Mai:2017bge,Hammer:2017uqm,Hammer:2017kms,Doring:2018xxx,Jackura:2018xnx,Dawid:2020uhn,Muller:2021uur,Hansen:2014eka,Hansen:2015zga,Blanton:2020gha,Briceno:2017tce,Briceno:2018aml,Hansen:2020zhy,Blanton:2020jnm,Blanton:2021mih}. Furthermore, these methods are applicable for
extracting three-body interaction information from lattice QCD data,
as demonstrated in Refs.~\cite{Meissner:2014dea,Horz:2019rrn,Blanton:2019vdk,Mai:2019fba,Culver:2019vvu,Fischer:2020jzp,Brett:2021wyd,Blanton:2021llb,Alexandru:2020xqf,Draper:2023boj,Hansen:2020otl}. 

The Efimov effect is a remarkable phenomenon in three-body physics \cite{Efimov:1970zz}. It provides a universal binding mechanism for three-body systems independent of the details of their interactions at short distances. In its simplest setting, it states that in a system of three identical bosons, when the interaction between two particles approaches the unitarity limit 
(that is when the pair scattering length $a$ is much larger than the range of the interaction $R$ and approaches infinity), a geometric spectrum of infinitely many three-body bound states emerges. The ratio of the binding energies ($E_{n}$) of these bound states follows the scaling law: 
 \begin{equation}
     \frac{E_{n}}{E_{n+1}}\approx 515.
	 \label{eq:Efimov scaling}
 \end{equation}
The exact unitary limit ($1/a \to 0, R\to 0$) is a theoretical construct and cannot be reached
experimentally (but it is possible to tune $1/a$ to zero in ultracold atoms using Feshbach resonances~\cite{Chin:2010crf}). However, the predictions of Efimov universality also apply for finite scattering length if $|a| \gg R$ (see Ref.~\cite{Braaten:2004rn} for a review). Three-body bound states are usually considered
Efimov states if they can be described by Efimov's universal
equation with small range corrections \cite{Efimov:1971zz,Braaten:2002sr}.
In particular, the ratio of subsequent Efimov states can deviate
significantly from the scaling law, Eq.~(\ref{eq:Efimov scaling}), close to the threshold if $|a|\gg R$ is finite \cite{Braaten:2004rn}.
Experimental evidence for the existence of an Efimov trimer was first  found in 2006 in a gas of ultracold Cs atoms~\cite{Kraemer_2006} by observing a unique three-body loss  signature \cite{Braaten:2004rn} that can be detected by varying the scattering length 
using the Feshbach resonance technique.
Since then, Efimov states have been observed for many bosonic and fermionic atoms as well as atomic mixtures (see Ref.~\cite{Naidon:2016dpf} for a review of these efforts).

The observation of Efimov physics in nuclear and particle physics systems is hindered by the lack of experimental control over the scattering length \cite{Hammer:2010kp}. Nevertheless, the triton and certain two-neutron halo nuclei can be considered approximate Efimov states, see, e.g., Refs.~\cite{Epelbaum:2008ga,Hammer:2017tjm,Hammer:2019poc}.
Moreover, the recent identification of new hadronic states, such as $X(3872)$ and $T^{+}_{cc}$, which are situated in the immediate vicinity of relevant two-body thresholds~\cite{Belle:2003nnu,LHCb:2021auc} 
opens up the potential to investigate the possible presence of Efimov states within three-body charmed hadron systems \cite{Braaten:2003he,Canham:2009zq,Hammer:2010kp,Valderrama:2018azi,MartinezTorres:2020hus}.

There have already been some results for three-charm systems using different methods~\cite{Canham:2009zq,Valderrama:2018sap,Wang:2020fuh,Luo:2021ggs,Bayar:2022bnc,Ortega:2024ecy,Zhu:2024hgm}. In particular, the emergence of the Efimov effect in the $D^*D^*D^*$ system has been explored in Ref.~\cite{Ortega:2024ecy}, under the assumption that a $D^*D^*$ molecule with quantum numbers $(I)J^P=(0)1^+$ exists as the heavy partner of the $T_{cc}^+$.
In this paper, we will examine the existence of the Efimov effect in the $D^*D^*D^*$ and $DD^*D^*$ systems using a systematic nonrelativistic EFT method, which is more convenient for treating all channels with different quantum numbers. 
We assume the existence of an isoscalar heavy spin partner of the $T_{cc}$, referred to as $T^*_{cc}$, which lies close to and below the $D^*D^*$ threshold, as predicted from the  existence of $T_{cc}(3875)$ using heavy quark spin symmetry in Refs.~\cite{Albaladejo:2021vln, Du:2021zzh}.\footnote{The hadronic and radiative decays of the $T_{cc}^*$ have been investigated in Refs.~\cite{Jia:2022qwr, Jia:2023hvc}. The existence of an isoscalar $D^*D^*$ bound state and its width were also investigated in Refs.~\cite{Molina:2010tx,Dai:2021vgf}.}
Both the $T_{cc}$ and $T^*_{cc}$ tetraquark states are characterized by the quantum numbers $(I,J)=(0,1)$, and they strongly couple to the $DD^*$ and $D^*D^*$ channels in $S$-wave, respectively. In Sec.~\ref{sec:NREFT}, we construct the scattering equations for different spin configurations ($T_{cc}D^*$ and $T^*_{cc}D^*$) within a pionless EFT, for all of which the total isospin is $I=1/2$. Section~\ref{sec:dis} will address the necessary conditions for the occurrence of the Efimov effect in the full $D^{*}D^*D^*$ system with all possible $(I,J)$ quantum numbers. A summary is given in the last section.

\section{$D^{*}D^{*}D^{*}$ and $DD^{*}D^{*}$ in NREFT\label{sec:NREFT}}

\subsection{Effective Lagrangian and propagators}

In an energy range very close to the threshold, the system can be described by a pionless EFT, which is dominated by contact interaction terms, as discussed in, e.g., Refs.~\cite{Bedaque:1998kg,Braaten:2003he,Wilbring:2016bda,Wilbring:2017fwy}. 
We consider that there is an isoscalar $T_{cc}$ near-threshold bound state in the $S$-wave $DD^*$ channel~\cite{LHCb:2021auc,LHCb:2021vvq} and an isoscalar $T^*_{cc}$ near-threshold bound state in the $S$-wave $D^*D^*$ channel as predicted in Refs.~\cite{Albaladejo:2021vln,Du:2021zzh}, and in this section we assume that there are no other near-threshold $S$-wave poles in two-body double-charm systems.\footnote{It was found in recent lattice QCD calculations that both the isovector $S$-wave $DD^*$~\cite{Meng:2024kkp} and $DD$~\cite{Shi:2025ogt} systems are repulsive.}

\begin{figure}[tb]
	\begin{center}
		\includegraphics[width=0.68\textwidth]{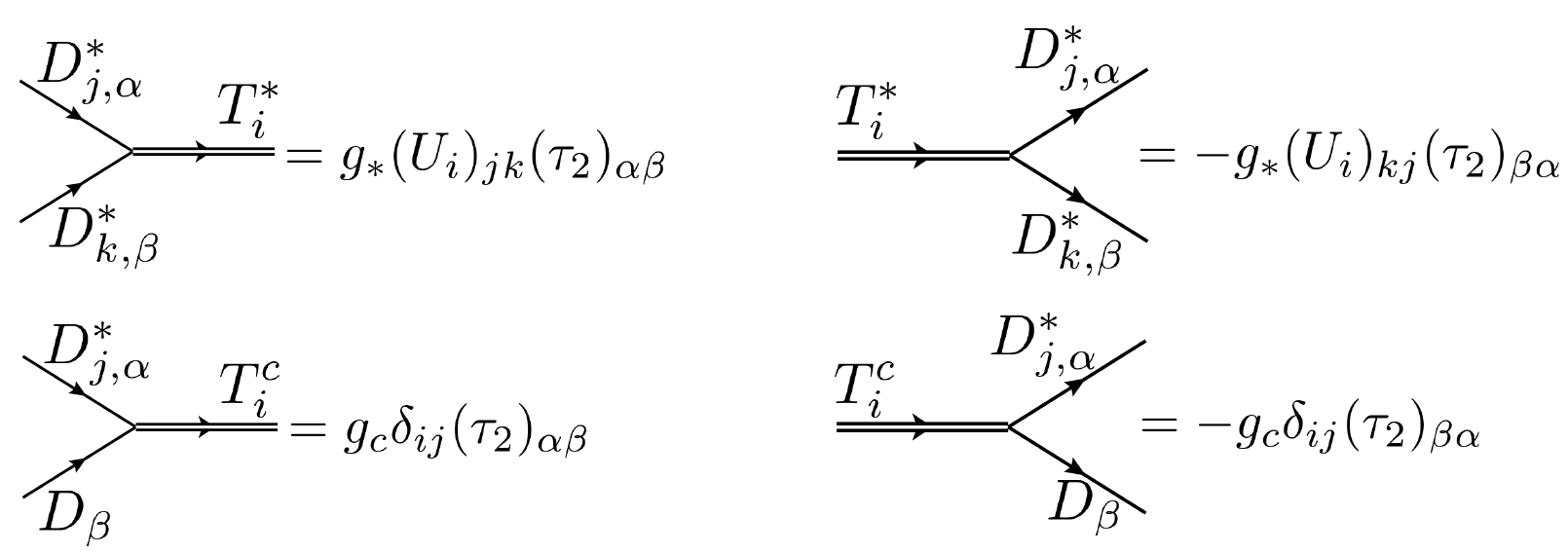}
	\end{center}
	\caption{\label{fig: feynrule}{Feynman rules for the dimer-particle couplings from the Lagrangian in Eq.~(\ref{eq:Lag}). }}
\end{figure}

The effective Lagrangian including the $T_{cc}$ and $T^*_{cc}$ dimer fields is given by
\begin{align}\label{eq:Lag}
    \mathcal{L} =&\, D_{i\alpha}^{*\dagger} \left(i\partial_{0} + \frac{\nabla^{2}}{2M_D}\right) D^*_{i\alpha}+D_{\alpha}^{\dagger} \left(i\partial_{0} + \frac{\nabla^{2}}{2M_{D^*}}\right) D_{\alpha} +T_{i}^{*\dagger}\Delta_{*} T_{i}^*+T_{i}^{c\dagger}\Delta_{c}T_{i}^c \nonumber\\
    &-g_{*}\left[T_{i}^{*\dagger}D^{*}_{j\alpha}(U_{i})_{jk}(i\tau_{2})_{\alpha\beta}D^{*}_{k\beta}+ \text{h.c.}\right] -g_{c}\left[T_{i}^{c\dagger}D^{*}_{j\alpha}\delta_{ij}(i\tau_{2})_{\alpha\beta}D_{\beta}+ \text{h.c.}\right],
\end{align}
where $D$ and $D^*$ denote the pseudoscalar and vector charm meson fields, respectively, while $T^*$ and $T^c$ represent the dimer fields that annihilate the $T_{cc}^*$ and $T_{cc}$ states, respectively. The lower-case Latin letters $i,j,k \in \{1,2,3\}$ serve as spin-1 indices, and the Greek lower-case letters $\alpha,\beta \in \{1, 2\}$ represent isospin-1/2 indices. The second Pauli matrix $\tau_2$ operates in the isospin space, while the $U_i$'s are the generators of the rotation group acting on the spin-1 representation, with matrix elements $(U_i)_{jk}=-i\epsilon_{ijk}$. 
The parameters $g_*$ and $g_c$ characterize the coupling strengths between the dimer fields and their constituent particles.
At leading order, $g_{*,c}$ and $\Delta_{*,c}$ are not independent parameters, and only the combinations $g_{*,c}^2/\Delta_{*,c}$ appear in physical observables.
The dimer-particle couplings are depicted in Fig.~\ref{fig: feynrule}.

The two-body scattering amplitude is proportional to the propagator of the dimer field. The dressed propagator $G_{A}$ is given by:
\begin{equation}
    iG_{A}(p_0,\mathbf{p})_{ij}=\frac{i\delta_{ij}}{\Delta_A+c_A \Sigma_A(p_0,\mathbf{p})},
\end{equation} 
where $A=*,c$ is the dimer index, and $\Sigma_{A}$ is the self-energy function. The spin and isospin operators between the dimer and meson fields contribute a Kronecker delta $\delta_{ij}$ in the numerator and a normalization constant $c_A$ in the denominator. For the propagator $G_{*}$, $c_*=4$, while for $G_{c}$, $c_c=2$.

The ultraviolet (UV) divergence in the self-energy can be regularized using the power divergence subtraction scheme with a hard scale $\Lambda_{\rm{PDS}}$~\cite{Kaplan:1998we,Kaplan:1998tg} as
\begin{align}
\Sigma_A(p_0,\mathbf{p}) = \frac{1}{2\pi}g_A^2S_{2, A}\mu_A\left[-\sqrt{-2\mu_A\left(p_0-\frac{p^2}{2M_{A}}\right)-i\varepsilon}+\Lambda_{\rm{PDS}}\right],
\end{align}
where $S_{2, A}$ represents the symmetry factor associated with the exchange of identical particles in $\Sigma_A$ (for the $D^*D^*$ and $DD^*$ dimers, $S_{2, *}=2$ and $S_{2, c}=1$, respectively),
and $p_0$ and $\mathbf{p}$ are the energy and three-momentum of the dimer, with $p$ denoting the magnitude of $\mathbf{p}$. 
Here, $\mu_A$ and $M_A$ denote the reduced mass and total mass of the constituent particles, respectively. Specifically, $\mu_c=M_D M_{D^*}/M_c$ with $M_c=M_D+M_{D^*}$ for the $DD^*$ system, and $\mu_*=M_{D^*} M_{D^*}/M_*$ with $M_*=2 M_{D^*}$ for the $D^*D^*$ system. 

The renormalized propagators for the $T_{cc}^*$ and $T_{cc}$ dimers are then obtained as follows:
\begin{align}   
	\begin{aligned}  
   & iG_{*}(p_0,\mathbf{p})_{ij}=iG_{*}(p_0,\mathbf{p})\delta_{ij} = -\frac{2\pi i}{g^2_{*}\mu_{*}S_{2,*}c_*}\frac{\delta_{ij}}{-1/a_{*}+\sqrt{-2\mu_{*}(p_{0}-\frac{\mathbf{p}^2}{2M_{*}}+i\varepsilon)}},\\
   & iG_{c}(p_0,\mathbf{p})_{ij}=iG_{c}(p_0,\mathbf{p})\delta_{ij} = -\frac{2\pi i}{g^2_{c}\mu_{c}S_{2,c}c_c}\frac{\delta_{ij}}{-1/a_{c}+\sqrt{-2\mu_{c}(p_{0}-\frac{\mathbf{p}^2}{2M_{c}}+i\varepsilon)}},
	\end{aligned}
\end{align}
The UV divergence in the self-energy has been absorbed into the scattering length through the relation
 \begin{equation}
     \frac{1}{a_{A}}=\frac{2\pi\Delta_{A}}{g_A^2\mu_AS_{2,A}c_A}+\Lambda_{\rm{PDS}}.
 \end{equation}
The value of $a_{A}$ determines the pole position of the two-body subsystem, with a positive $a$ corresponding to a bound state and a negative $a$ corresponding to a virtual state.\footnote{For a repulsive potential, the scattering length has the same sign as that in the case with one bound state, but the magnitude is much smaller such that the pole is very far below 2-body threshold and thus beyond the applicable range of theory.} 
Furthermore, the wave function renormalization constants $Z_{*}$ and $Z_c$, which are obtained from the residue at the pole are given by
 \begin{equation}
     Z_*=\frac{2\pi\gamma_*}{g_*^2\mu^2_*S_{2,*}c_*},\quad Z_c=\frac{2\pi\gamma_c}{g_c^2\mu^2_cS_{2,c}c_c},
 \end{equation}
where $\gamma_A = \pm\sqrt{2 \mu_A |E_B^A|}$ ($A=*,c$ for $T_{cc}^*$ and $T_{cc}$, respectively) denotes the binding momentum with the positive (negative) sign for a bound (virtual) state pole and $E_B^A$ the corresponding binding energy.

\subsection{Three-body scattering equations\label{sec:eq}}

Using the particle-dimer Lagrangian, we can derive the scattering equation for the three-body system.
Our calculations are performed in the center-of-mass (c.m.) frame of the corresponding spectator-dimer system, such as the $D^{(*)}T_{A}$ system. The total energy $E$ is expressed as:
\begin{equation}
    E=\frac{q^2}{2M_A}+\frac{q^2}{2M_\text{sp}}-\frac{\gamma_A^2}{2\mu_A},
\end{equation}
where $M_\text{sp}=M_{D}(M_{D^*})$ is the mass of the spectator, $M_{A}=M_{*}(M_{c})$ is the total mass of the two particles in the two-body subsystem, and $q$ is the momentum of the spectator.

We analyze the scattering equation to investigate possible bound states. 
The absence of the strong cutoff dependence of observables (e.g., the scattering amplitudes) in our calculations indicates that there are no bound states in the corresponding channel~\cite{Bedaque:1998km,Canham:2009zq,Wilbring:2017fwy}. Conversely, the emergence of three-body bound states signals the Efimov effect. 
In such cases, the integral equation, which includes only two-body interactions, exhibits significant cutoff dependence, necessitating the inclusion of three-body contact interactions for renormalization~\cite{Bedaque:1998kg,Bedaque:1998km}. 
For the scattering equation of three identical particles in various partial wave channels, the existence of the Efimov effect depends on the coefficients in the scattering equation of the system under investigation~\cite{Braaten:2004rn,Wilbring:2016bda}. These coefficients include contributions from the Clebsch-Gordan (CG) coefficients of (iso)spin coupling and the symmetry factor of identical particles. 
A systematic analysis, for which the dimers are not restricted to the $T_{cc}^*$ and $T_{cc}$, will be presented in the subsequent section.

\subsection{$T^*_{cc}D^*$ Scattering}

\begin{figure*}[tbh]
    \centering
    \includegraphics[width=0.9\textwidth]{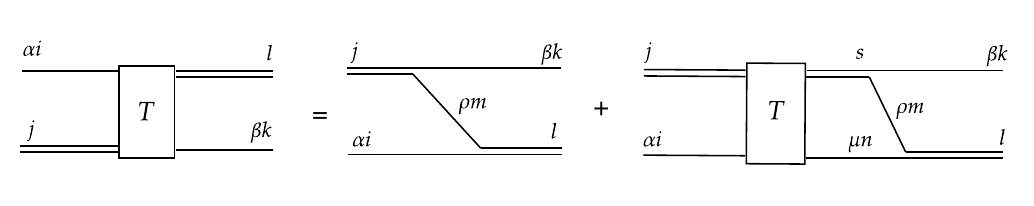}
    \caption{Integral equation for the $T^*_{cc}D^*$ scattering amplitude $T$ with incoming isospin index $\alpha$ and spin indices $i,j$, and outgoing indices $\beta$ and $k,l$. Here, the single solid line represents the $D^*$ meson while the double solid line denotes the $D^*D^*$ dimer $T_{cc}^*$.\label{fig:3Dstar}}
\end{figure*}

\begin{figure}[tb]
    \centering
    \includegraphics[width=0.3\textwidth]{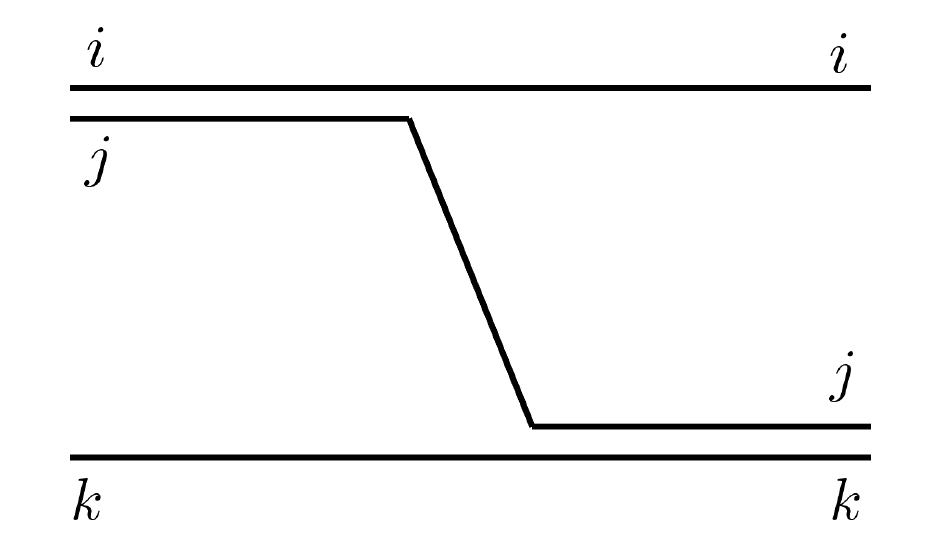}
    \caption{Diagram for particle-dimer scattering via the particle exchange. Here $i,j$ and $k$ refer to particles.}
    \label{fig:S3}
\end{figure}

We begin with the single-channel $T^*_{cc}D^*$ scattering, whose amplitude $T$ satisfies the integral equation shown in Fig.~\ref{fig:3Dstar}. Before proceeding further, let us present the independent tree-level amplitudes for the $T^{*(c)}D^{(*)}$ systems. Using the vertex factors shown in Fig.~\ref{fig: feynrule}, we have
\begin{align}\label{eq: tree_amp} 
	\begin{aligned}
	i {\cal M}_{T^*D^*\to T^*D^*}^{j i \alpha \to l k \beta} (E, \mathbf{k}, \mathbf{p}) &={} \frac{-i g_*^{2}S_{3,1}(U_l U_j)_{ik}\delta_{\beta\alpha}}{E-\frac{k^2}{2M_{D^*}}-\frac{p^2}{2M_{D^*}}-\frac{(\mathbf{k}+\mathbf{p})^2}{2M_{D^*}}+i\eps},\\
	i {\cal M}_{T^*D\to T_cD^*}^{i \alpha \to j k \beta} (E, \mathbf{k}, \mathbf{p}) &={} \frac{-i g_*g_cS_{3,2}(U_i)_{k j}\delta_{\beta\alpha}}{E-\frac{k^2}{2M_{D}}-\frac{p^2}{2M_{D^*}}-\frac{(\mathbf{k}+\mathbf{p})^2}{2M_{D^*}}+i\eps},\\
	i {\cal M}_{T_cD^*\to T_c D^*}^{j i \alpha \to l k \beta} (E, \mathbf{k}, \mathbf{p}) &={} \frac{-i g_c^2S_{3,3}\delta_{i l}\delta_{k j}\delta_{\beta\alpha}}{E-\frac{k^2}{2M_{D^*}}-\frac{p^2}{2M_{D^*}}-\frac{(\mathbf{k}+\mathbf{p})^2}{2M_{D}}+i\eps},\\
	i {\cal M}_{T_c D^*\to T^*D}^{j i \alpha \to k \beta} (E, \mathbf{k}, \mathbf{p}) &={} \frac{-i g_*g_cS_{3,4}(U_k)_{ji}\delta_{\beta\alpha}}{E-\frac{k^2}{2M_{D^*}}-\frac{p^2}{2M_{D}}-\frac{(\mathbf{k}+\mathbf{p})^2}{2M_{D^*}}+i\eps},
	\end{aligned}
\end{align}
where $\mathbf{k},\mathbf{p}$ represent the outgoing and incoming 3-momenta, respectively. $S_{3,i}$ is the symmetry factor for each one-boson-exchange diagram, with $S_{3,1}=4$, $S_{3,2}=S_{3,4}=2$ and $S_{3,3}=1$. These symmetry factors due to the exchange of identical particles are given by (see Fig.~\ref{fig:S3})
\begin{align}
S_{ijk} &= \left\{ 
  \begin{array}{ll}
  4 & \text{if } i=j=k, \\
  2 & \text{if } i\neq j=k \text{ or } i=j\neq k, \\
  1 & \text{if } i=k\neq j.
  \end{array}
\right.
\end{align} 

The integral equation for the $T^*_{cc}D^*$ scattering is given by
\begin{align}
	&t_{j i\alpha}^{l k \beta} (E, \mathbf{k}, \mathbf{p}) ={} {\cal M}_{T^*D^*\to T^*D^*}^{j i \alpha \to l k \beta} (E, \mathbf{k}, \mathbf{p}) \nonumber\\
	&\phantom{xx} + \: \int \frac{d^4 q}{(2\pi)^4} \:
	\frac{i{\cal M}_{T^*D^*\to T^*D^*}^{s n \mu \to l k \beta} (E, \mathbf{q}, \mathbf{p})}{-q_0 - \frac{q^2}{2M_{D^*}} + i \varepsilon}\times  \frac{2\pi}{g_*^2 \mu_*}\frac1{S_{2,*} c_*}\frac{t_{j i \alpha}^{s n \mu} (E, \mathbf{k}, \mathbf{q})}{-\gamma_* + \sqrt{-2\mu_* \left(E + q_0 - \frac{q^2}{2M_*} \right)
			- i\varepsilon}},
\end{align}
where $t_{j i \alpha}^{l k \beta}$ is the scattering amplitude that includes the complete
spin-isospin structure.
After integrating over the $q_0$ component and multiplying by the wave function
renormalization constant, we obtain
\begin{align}
	& T_{j i\alpha}^{l k \beta} (E, \mathbf{k}, \mathbf{p}) ={} -\frac{\pi \gamma_*}{\mu_*^2}\frac{2S_{3,1}}{S_{2,*}c_*} \frac{(U_l U_j)_{ik}\:\delta_{\beta\alpha}}{E - \frac{k^2}{2M_{D^*}} - \frac{p^2}{2M_{D^*}} - \frac{(\mathbf{k} + \mathbf{p})^2}{2M_{D^*}} + i\varepsilon} \nonumber\\
	&\quad - \frac{\pi}{\mu_*}\frac{2S_{3,1}}{S_{2,*}c_*} \int \frac{d^3 q}{(2\pi)^3} \:
	\frac{T_{j i\alpha}^{s n \mu} (E, \mathbf{k}, \mathbf{q})}{-\gamma_* + \sqrt{-2\mu_* \left(E - \frac{q^2}{2M_{D^*}} - \frac{q^2}{2M_*} \right) - i\varepsilon}}  \frac{(U_l U_s)_{nk}\:\delta_{\beta\mu}}{E - \frac{q^2}{2M_{D^*}} - \frac{p^2}{2M_{D^*}} - \frac{(\mathbf{q} + \mathbf{p})^2}{2M_{D^*}} + i\varepsilon} \:,
\end{align}
with $T_{j i \alpha}^{l k \beta} \equiv Z_* \: t_{j i \alpha}^{l k \beta}$.

$T_{j i \alpha}^{l k \beta}$ can be projected onto a general partial wave,
\begin{align}
	\frac{1}{2} \int_{-1}^{+1} d\cos \theta \: P_L (\cos \theta) \: T(E, \mathbf{k}, \mathbf{p}) = T_{(L)} (E,k,p) \:,
\end{align}
where $\theta$ is the angle between $\textbf{k}$ and $\textbf{p}$, $P_L(\cos \theta)$ is the Legendre polynomial of order $L$. 
For $L=0$,
we obtain the integral equation for the $S$-wave
$T^*_{cc}D^*$ scattering amplitude
\begin{align}\label{eq: TstarDstar}
	T_{(0) \, j i\alpha}^{\quad l k \beta} (E,k,p) =&{} -\frac{\pi \gamma_*}{\mu_*^2}\frac{2S_{3,1}}{S_{2,*}c_*} V^S_1(k,p) \boxed{(U_l U_j)_{ik}\:\delta_{\beta\alpha}} \nonumber\\& - \frac{1}{\pi\mu_*}\frac{S_{3,1}}{S_{2,*}c_*} \int_0^{\Lambda} dq \frac{q^2 \:
		V^S_1(q,p) \: T_{(0) \, j i \alpha}^{\quad s n \mu} (E,k,q) \boxed{(U_l U_s)_{nk}\:\delta_{\beta\mu}}}{-\gamma_* + \sqrt{-2\mu_* \left(E - \frac{q^2}{2M_{D^*}} - \frac{q^2}{2M_*} \right)
			- i\varepsilon}} \nonumber\\
	\equiv&\, {\cal M}_0\:\boxed{C_{0\;j i \alpha}^{\,\,\, lk \beta}}+\int_0^{\Lambda} dq\: {\cal M}_1 T_{(0) \, j i \alpha}^{\quad sn \mu} (E,k,q)\:\boxed{C_{0\; sn \mu}^{\,\,\,lk \beta}},
\end{align}
where $\Lambda$ is the UV cutoff discussed above. 
The last equality defines the amplitudes ${\cal M}_0$, ${\cal M}_1$ and the coefficients $C_{0\;j i \alpha}^{\,\,\, lk \beta}$ and $C_{0\; sn \mu}^{\,\,\,lk \beta}$; for the latter we use boxed notations to make the corresponding definitions more transparent (similar notations will be used later).
The $S$-wave potential $V^S$ is given by
\begin{equation}
	V^S_{1}(k,q)=-\frac{M_{D^*}}{k p}Q_0\left(-\frac{M_{D^*}}{k p}\left(E-\frac{k^2}{2\mu_*}-\frac{p^2}{2\mu_*}\right)-i\eps \right).
\end{equation}
The logarithmic function $Q_0$ originates from the one-meson exchange contributions, whose
$S$-wave projection leads to integrals of the form
\begin{align}
	Q_0(\beta)\equiv \frac12\int_{-1}^{+1} dx \frac{P_{0} (x)}{x+\beta} = \frac12\ln \left(\frac{\beta+1}{\beta-1} \right).
\end{align}

For the $S$-wave $T^*_{cc} D^*$ system, all possible quantum numbers are $J^P=0^-,1^-$ and $2^-$. Following Ref.~\cite{Wilbring:2016bda},
we project out the desired channel with a given isospin $I$ and angular momentum $J$ by evaluating: 
\begin{align}\label{eq: projection}
	T_{(0)}^{I, J} \equiv \frac{1}{(2J+1)(2I+1)} \sum_{\substack{\tilde{m}\tilde{\eta}, \tilde{n}\tilde{\lambda}}} {\cal O}_{\;\tilde{n}\tilde{\lambda},\tilde{j}\tilde{\beta}}^{\dagger} \: T_{(0) \, \tilde{i}\tilde{\alpha}}^{\quad \tilde{j}\tilde{\beta}} \: {\cal O}^{}_{\tilde{m}\tilde{\eta},\tilde{i}\tilde{\alpha}}\:,
\end{align}
where we use the tilded indices to represent both spin and isospin (if any) indices of the given operators. $\tilde{i}\tilde{\alpha}$ and $\tilde{j}\tilde{\beta}$ are the indices for initial and final state spectator-dimer pairs. To be more specific, $\tilde{i}$ ($\tilde{j}$) includes both the spin and isospin indices for the initial (final) $D^*$, while $\tilde{\alpha}$, $\tilde{\beta}$ are the spin indices for the isoscalar $T^*_{cc}$. $\tilde{n}\tilde{\lambda}$ and $\tilde{m}\tilde{\eta}$ represent the indices in the corresponding $(I,J)$ channel after projection.
It should be noted that for elastic scattering, the initial and final states are identical, requiring $\tilde{\eta}=\tilde{\lambda}$ and $\tilde{m}=\tilde{n}$.
The projection operators for the scalar, vector and tensor amplitudes are
\begin{align}\label{eq: spin_operator}
	{\cal O}_{ j i}^{J=0}\left(1\otimes 1\to 0\right)=\:&\frac{-1}{\sqrt{3}}\delta_{ij}\:,\notag\\
	{\cal O}_{ \ell,m n}^{J=1}\left(1\otimes 1\to 1\right)=\:&\frac{-1}{\sqrt{2}}(U_\ell)_{m n}\:, \\
	{\cal O}_{\ell k, m n}^{J=2}\left(1\otimes 1\to 2\right)=\:&\frac12[\delta_{\ell m}\delta_{k n}+\delta_{\ell n}\delta_{k m}-\frac23\delta_{\ell k}\delta_{m n}]\:, \notag
\end{align}
and the corresponding isospin projection operator is given by 
\begin{equation}
    {\cal O}_{\alpha, \beta}^{I=1/2}\left(\frac12\otimes 0\to \frac12\right)=\:\delta_{\alpha\beta}.
\end{equation}

Applying the above projections
to the integral equation of the $S$-wave $T^*_{cc} D^*$ scattering, Eq.~\eqref{eq: TstarDstar}, one gets
\begin{align}\label{eq: TstarDstar_pw1}
	T_{(0)}^{I=\frac{1}{2}, J=0} \equiv&\, \frac{1}{2} \sum_{\substack{\eta\lambda}} \bigg({\cal O}_{\lambda, \beta}^{I=1/2}{\cal O}_{l k}^{J=0}\bigg)^\dagger  \: T_{(0) \, ji \alpha}^{\quad l k \beta} \bigg({\cal O}_{\eta,\alpha}^{I=1/2}{\cal O}_{ji}^{J=0}\bigg)\notag\\
	=&\, \frac1{2}\frac13\left[\delta_{kl}(U_l U_j)_{ik}\delta_{ji}\right]\Bigg[\sum_{\substack{\eta\lambda}}\delta_{\beta\lambda}\delta_{\beta\alpha}\delta_{\eta\alpha}\Bigg]{\cal M}_0 \notag \\
	& +\frac{1}{2}\int_0^{\Lambda} dq\: {\cal M}_1 \sum_{\substack{\eta\lambda}} \bigg(\frac{-1}{\sqrt{3}}\delta_{lk}\delta_{\beta\lambda}\bigg)  C_{0\; s n\mu}^{\,\,\, l k \beta} T_{(0) \, j i \alpha}^{\quad sn \mu} \bigg({\cal O}_{\eta,\alpha}^{I=1/2}{\cal O}_{ji}^{J=0}\bigg)\notag\\
	=&\,\frac1{2}\frac13\times(-6)\times2\: {\cal M}_0+\int_0^{\Lambda} dq\: {\cal M}_1(-2)\times\frac1{2} \sum_{\substack{\eta\lambda}} \bigg({\cal O}_{\lambda, \mu}^{I=1/2}{\cal O}_{s n}^{J=0}\bigg)^\dagger T_{(0) \, ji \alpha}^{\quad s n \mu}\bigg({\cal O}_{\eta,\alpha}^{I=1/2}{\cal O}_{ji}^{J=0}\bigg)\notag\\
	=& -2\:{\cal M}_0-2\int_0^{\Lambda} dq\: {\cal M}_1 T_{(0)}^{I=\frac{1}{2}, J=0},
\end{align}
with two scalar amplitudes ${\cal M}_{0}$ and ${\cal M}_{1}$, which have been defined by the last equality in Eq.~\eqref{eq: TstarDstar}.
Similarly, applying the spin-isospin projections for other partial waves to Eq.~\eqref{eq: TstarDstar} yields the corresponding integral equations for each partial wave amplitude, all of which can be written in the following general form,
\begin{align}\label{eq: TstarDstar_pw}
	T_{(0)}^{I, J}(E,k,p) = C_0^IC_0^J\:{\cal M}_{0}(k,p)+\int_0^{\Lambda} dq\: C_0^IC_0^J {\cal M}_{1}(p,q) T_{(0)}^{I, J}(E,k,q).
\end{align}
The spin-isospin coefficients are given in Table~\ref{Tab: coeff-TstarDstar}.
\begin{table}[tb]
	\centering
	\renewcommand\arraystretch{1.5}
	\caption{Coefficients of the partial-wave projected integral equation for the $S$-wave $T^*_{cc} D^*$ scattering.\label{Tab: coeff-TstarDstar}}
	\begin{tabular}{l c c c}
		\hline
		\hline
		Channel & $(I,J)=(1/2,0)$ & $(I,J)=(1/2,1)$ & $(I,J)=(1/2,2)$ \\
		\hline
		$C_0^IC_0^J$  & $1\times(-2)$  & $1\times 1$	&	$1\times1$ \\
		\hline
		\hline
	\end{tabular}
\end{table}

The dimer-particle amplitude $T_{(0)}^{I,J}(E,k,p)$, obtained from the integral equations above, contains all physical information relevant to the three-meson system with specified quantum numbers. These include, for instance, the phase shifts of $T^{(*)}$-$D^{(*)}$ scattering and the possible existence of three-body bound states. After discretizing the integral in Eq.~\eqref{eq: TstarDstar_pw}, the integral equation can be rewritten as an eigenvalue problem expressed as follows:
\begin{equation}
    (1-C_0^I C_0^J \tilde{\cal M}_1) \tilde{T}_{(0)}^{I, J}=C_0^I C_0^J \tilde{\cal M}_0,
\end{equation}
where $\tilde{}$ denotes the discretized form (vector or matrix) of the quantity evaluated on the chosen integral mesh.
To explore the existence of such bound states, one can directly identify the poles of the amplitude, which are given as solutions of the following equation:
\begin{equation}
    \det(1-\tilde{V}\tilde{G}) \equiv \det(1-C_0^I C_0^J \tilde{\cal M}_1)=0,
\end{equation}
which is exactly analogous to the strategy used in searching for poles of the two-body scattering amplitude solved from the Lippmann-Schwinger equation.

\begin{figure}[tb]
    \centering
    \includegraphics[width=0.5\linewidth]{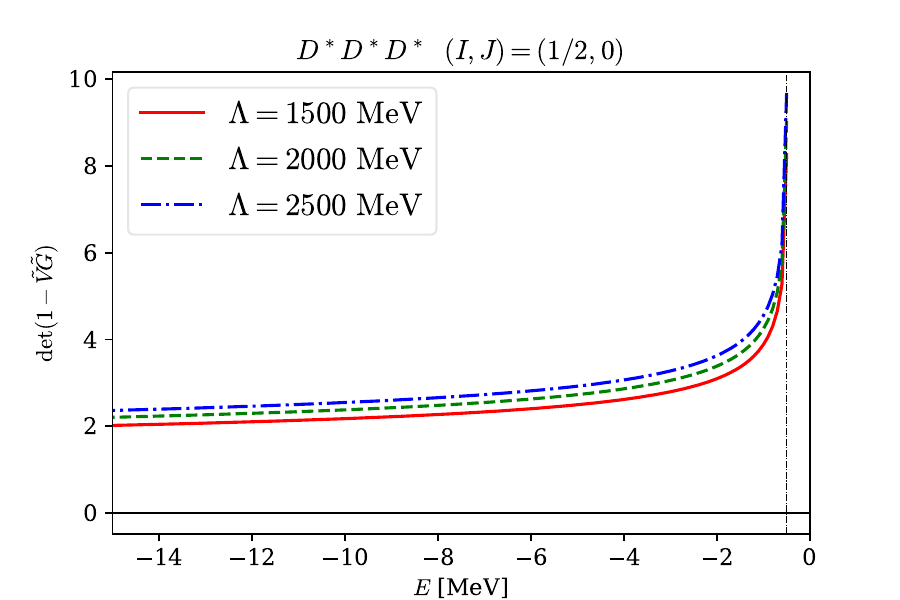}\hfill
    \includegraphics[width=0.5\linewidth]{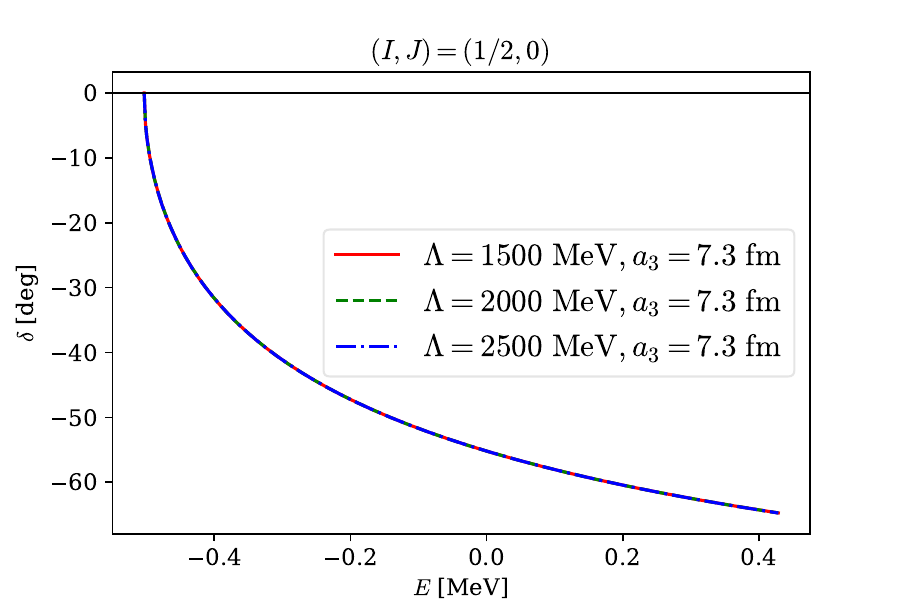}
	\includegraphics[width=0.5\linewidth]{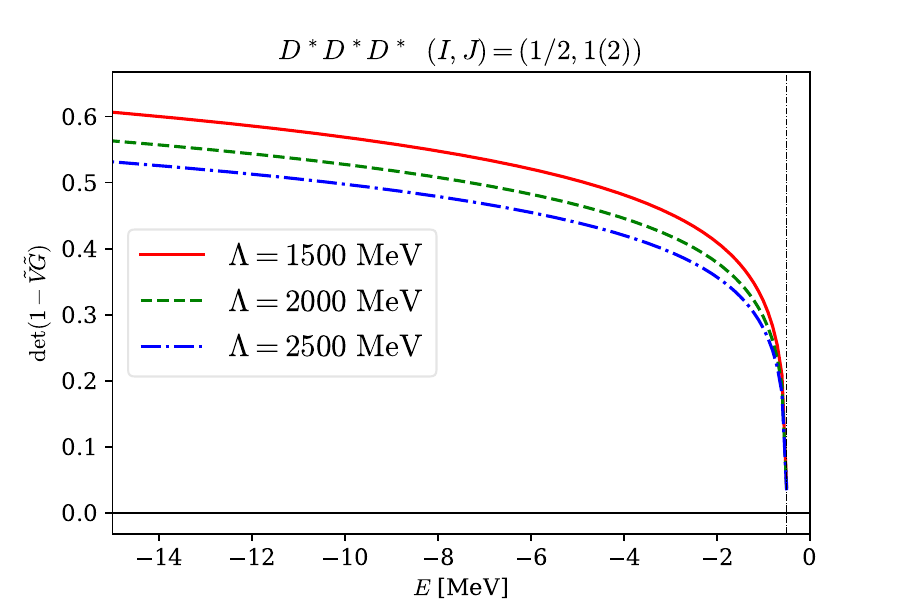}\hfill
    \includegraphics[width=0.5\linewidth]{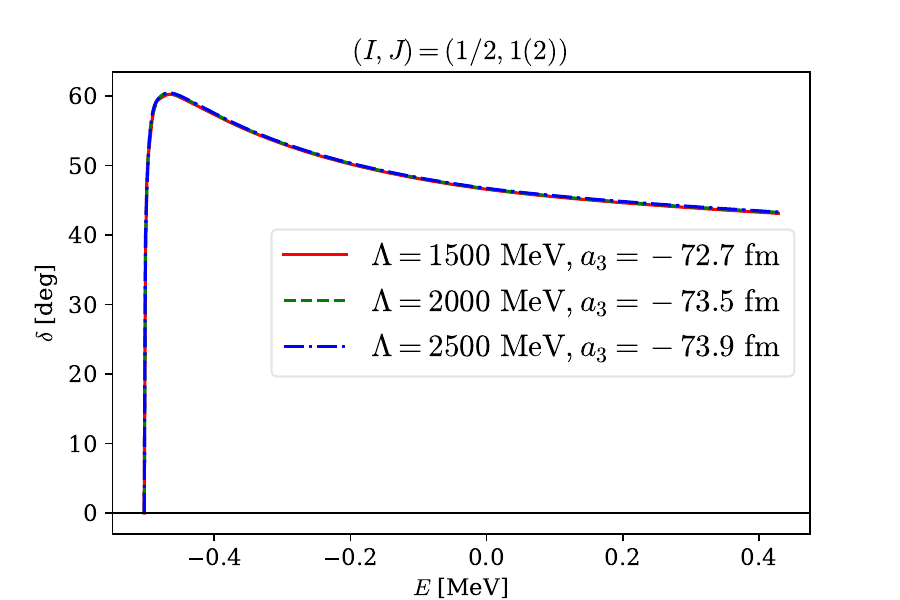}
    \caption{Left panel: determinant $\operatorname{det}(1-\tilde{V}\tilde{G})$ of the $T^*_{cc}D^*$ scattering equation for the channel with quantum numbers $(I,J)=(1/2,0)$ and $(I,J)=(1/2,1(2))$. Right panel: the corresponding $S$-wave $T_{cc}^*D^*$ scattering phase shifts, for which curves with different cutoffs are almost indistinguishable. The vertical lines denote the $D^*T^*_{cc}$ two-body threshold.
	\label{fig:detD0121}} 
\end{figure}

Our results demonstrate that all partial waves exhibit weak cutoff dependence, as shown in the left panel of Fig.~\ref{fig:detD0121}. Note that the only theoretical inputs are the binding energies of the two $(I,J)=(0,1)$ dimers, corresponding to physical $T_{cc}^*$ and $T_{cc}$ states, given by $E_B^*=-0.50$~MeV and $E_B^c=-0.36$~MeV~\cite{Du:2021zzh}, respectively.
The absence of trimer poles is evident across a wide range of $\Lambda$ values from $1.5$ to $2.5$ GeV.\footnote{
A different conclusion was reached in Ref.~\cite{Ortega:2024ecy} due to inconsistent prefactors arising from the partial wave projection compared to our work. There the existence of a three-body bound state with quantum numbers $(I,J)=(1/2,0)$ in the $D^*D^*D^*$ system was claimed, using a coefficient equivalent to $C_0^{J=0}=4$ in our notation. However, in our case, 
$C_0^{J=0}=-2$. In fact, as this coefficient changes from $4$ to $-2$, the three-body system undergoes a Berezinskii-Kosterlitz-Thouless-like phase transition \cite{Kaplan:2009kr}, leading to the disappearance of the bound state.}  

In the right panel of Fig.~\ref{fig:detD0121}, we have plotted the corresponding scattering phase shifts of $D^*T^*_{cc}$. One sees that the phase shifts remain almost unchanged when altering the cutoff $\Lambda$. In the plot, the $D^*T_{cc}^*$ $S$-wave scattering length $a_3$ is also labeled, whose definition is as follows:  
\begin{equation}\label{eq:phase}
    a^{(I,J)}_3 \equiv -\frac{\mu_{3}}{2\pi}T^{I,J}_{(0)}(k=p=0),
\end{equation}
where $\mu_{3}={M_{T^*_{cc}}M_{D^*}}/{(M_{T^*_{cc}}+M_{D^*})}$ is the reduced mass of $D^*T^*_{cc}$. The $S$-wave phase shifts for $D^*T^*_{cc}$ system are calculated as
\begin{equation}
    \delta^{(I,J)}=\mathrm{arccot}\left(\frac{2\pi}{\mu_{3}}\frac{1}{k}\,\mathrm{Re}\frac{1}{T^{(I,J)}_{(0)}(E,k,k)}\right).
\end{equation}

\begin{figure}[t]
    \centering
    \includegraphics[width=0.5\linewidth]{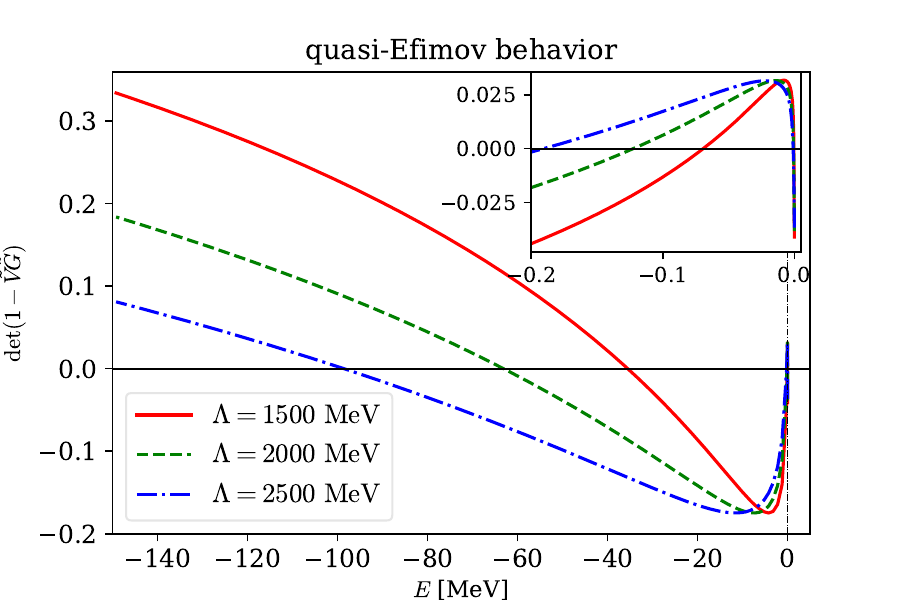}\hfill
    \includegraphics[width=0.5\linewidth]{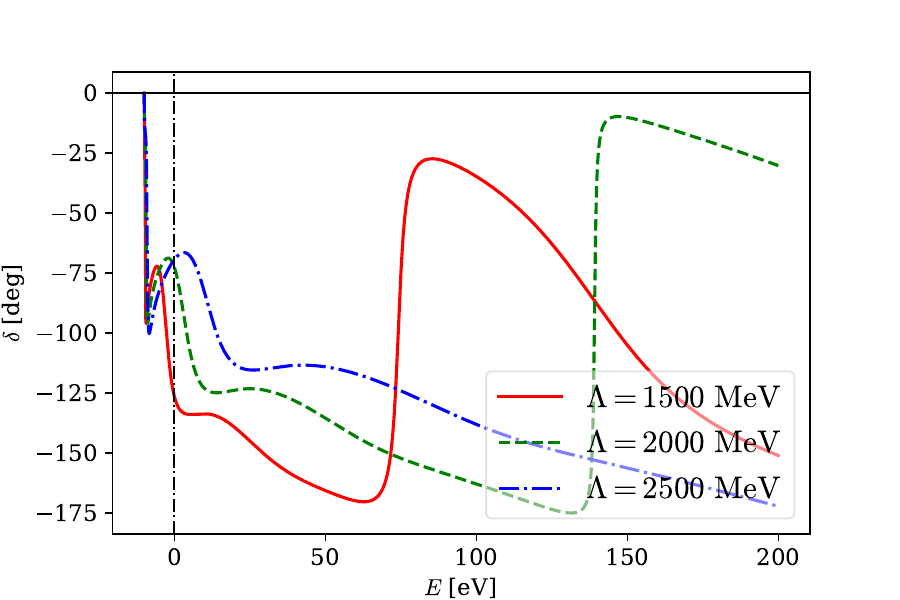}
    \caption{Left panel: determinant $\operatorname{det}(1-\tilde{V}\tilde{G})$ of the scattering equation for three identical scalar bosons with mass $m=M_{D^*}$, illustrating the 
    Efimov behavior. The two-body binding energy is set to 10~eV, and zero-crossings of the determinant indicate three-body bound states.
	The vertical line denotes the three-body threshold which almost equals to particle-dimer threshold in this binding energy. The inset magnifies the region very near threshold. Right panel: the particle-dimer scattering phase shift. Different from the results in Fig.~\ref{fig:detD0121}, the phase shift is very sensitive to $\Lambda$.
	} \label{fig:detD3a}
\end{figure}

To illustrate how $\operatorname{det}(1-\tilde{V}\tilde{G})$ and the phase shift would behave if an Efimov state exists (the scattering equation for such a system can be found in Refs.~\cite{Bedaque:1998kg,Bedaque:1998km}), in Fig.~\ref{fig:detD3a} we present a characteristic 
Efimov scenario with three identical scalar bosons of mass $M_{D^*}$ and a two-body binding energy of 10~eV, the value of which is only for illustration. 
Notice that no three-body contact term has been included to absorb the cutoff dependence (or in other words, the three-body contact term is set to zero).
From the left panel of Fig.~\ref{fig:detD3a}, one sees that tuning the two-body binding energy to 10~eV reveals the existence of at least three three-body bound states.
The right panel shows that the particle-dimer $S$-wave phase shift changes drastically within a small energy range due to the presence of multiple near-threshold bound state poles. One also notices that the phase shift is very sensitive to the cutoff, in contrast to the case without any Efimov state in Fig.~\ref{fig:detD0121}.
The cutoff dependence in both the phase shift and the three-body bound state position can be removed once a three-body contact term is included.

\subsection{$T^*_{cc} D$-$T_{cc} D^*$ coupled-channel scattering}
\begin{figure*}[t]
    \centering
    \includegraphics[width=0.9\textwidth]{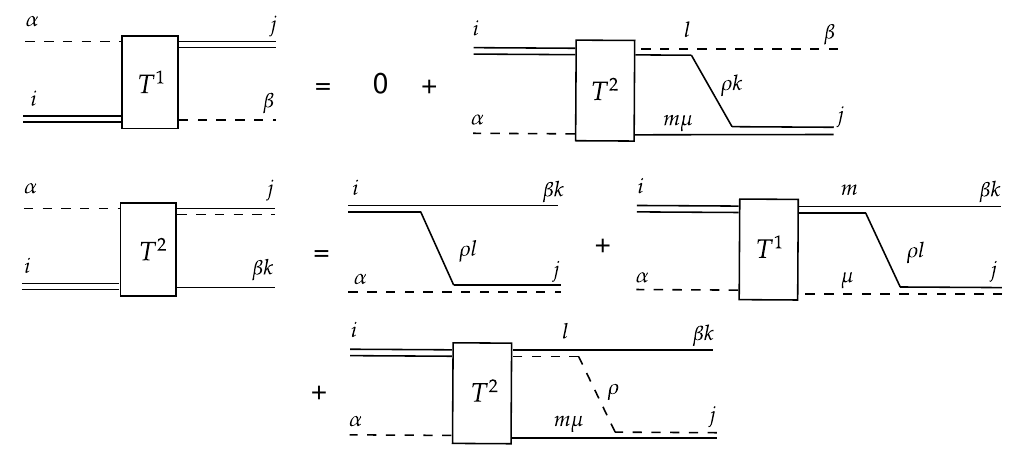}
    \caption{Integral equations for the coupled-channel ($T^*_{cc} D$ and $T_{cc} D^*$) scattering amplitudes. Here, the single solid (dashed) line represents $D^*$ ($D$) meson, and the double solid (solid-dashed) line represents the $D^*D^*$ ($DD^*$) dimer $T_{cc}^*$ ($T_{cc}$).
	\label{fig:2Dstar}}
\end{figure*}

Next, we consider the $S$-wave $T^*_{cc} D$-$T_{cc} D^*$ scattering.
For total spin $J=1$, we must solve coupled-channel integral equations since transitions between $T^*_{cc} D$ and $T_{cc} D^*$ are allowed, as illustrated in Fig.~\ref{fig:2Dstar}. In contrast, for $J=0$ and $J=2$, only the single-channel $T_{cc} D^*$ scattering occurs.

The coupled-channel integral equations for the $T^*_{cc} D\to T^*_{cc} D$ scattering amplitude $T_1$ and the $T^*_{cc} D\to T_{cc} D^*$ scattering amplitude $T_2$ are
\begin{align}\label{eq: TstarD-1}
	(T_1)_{(0) \, i \alpha}^{\quad \, j \beta} (E,k,p) &= -\frac{1}{\pi\mu_*}\frac{S_{3,4}}{S_{2,c}c_c} \int_0^{\Lambda} dq \sqrt{\frac{\gamma_*S_{2,c}c_c}{\gamma_cS_{2,*}c_*}} \frac{q^2 \:
		V_4^S(q,p)(T_2)_{(0) \, i \alpha}^{\quad l m \mu} (E,k,q)\: \boxed{(U_j)_{l m} \delta_{\beta\mu}}
	}{-\gamma_c + \sqrt{-2\mu_c \left(E - \frac{q^2}{2M_{D^*}} - \frac{q^2}{2M_c} \right)
			- i\varepsilon}} \notag\\
	&\equiv +\int_0^{\Lambda} dq\: {\cal M}_{12} (T_2)_{(0) \, i\alpha}^{\quad lm \mu} (E,k,q)\:\boxed{C_{1\; l m\mu}^{\,\,\, j \beta}}\:,
\end{align}
and
\begin{align}\label{eq: TstarD-2}
	(T_2)_{(0) \, i\alpha}^{\quad j k \beta} (E,k,p) =& -\frac{\pi \sqrt{\gamma_*\gamma_c}}{\mu_*\mu_c}\frac{2S_{3,2}}{\sqrt{S_{2,*}c_*S_{2,c}c_c}} V^S_2(k,p)\:\boxed{(U_i)_{k j}\delta_{\beta\alpha}} \notag \\
	&-\frac{1}{\pi\mu_c}\frac{S_{3,2}}{S_{2,*}c_*} \int_0^{\Lambda} dq \sqrt{\frac{\gamma_cS_{2,*}c_*}{\gamma_*S_{2,c}c_c}} \frac{q^2 \:
	V_2^S(q,p)(T_1)_{(0) \, i \alpha}^{\quad m \mu} (E,k,q)\:\boxed{(U_m)_{k j}\delta_{\beta\mu}}
	}{-\gamma_* + \sqrt{-2\mu_* \left(E - \frac{q^2}{2M_{D}} - \frac{q^2}{2M_*} \right)
			- i\varepsilon}} \nonumber\\
	&\phantom{} - \frac{1}{\pi\mu_c}\frac{S_{3,3}}{S_{2,c}c_c} \int_0^{\Lambda} dq \frac{q^2 \: V^S_3(q,p) \: (T_2)_{(0) \, i \alpha}^{\quad l m\mu} (E,k,q) \:\boxed{\delta_{k l}\delta_{m j}\delta_{\beta\mu}}}{-\gamma_c + \sqrt{-2\mu_c \left(E - \frac{q^2}{2M_{D^*}} - \frac{q^2}{2M_c} \right)
	- i\varepsilon}} \nonumber\\
	\equiv&\, {\cal M}_{20}\:\boxed{C_{2\; i \alpha}^{\,\,\, jk \beta}}+\int_0^{\Lambda} dq\: {\cal M}_{21} (T_1)_{(0) \, i \alpha}^{\quad m \mu} (E,k,q)\:\boxed{C_{2\; m \mu}^{\,\,\,jk \beta}}\notag\\
	&\qquad \qquad\quad +\int_0^{\Lambda} dq\: {\cal M}_{22} (T_2)_{(0) \, i \alpha}^{\quad l m\mu} (E,k,q)\:\boxed{C_{3\; lm \mu}^{\,\,\,jk \beta}}\:,
\end{align}
where the $S$-wave projection and wave function renormalization factors have been applied, and the relevant $S$-wave potentials $V^S$ are given by
\begin{align}
	&V^S_{2}(k,q)=-\frac{M_{D^*}}{k p}Q_0\left(-\frac{M_{D^*}}{k p}\left(E-\frac{k^2}{2\mu}-\frac{p^2}{2\mu_*}\right)-i\eps \right)\:,\notag\\
	&V^S_{3}(k,q)=-\frac{M_{D}}{k p}Q_0\left(-\frac{M_{D}}{k p}\left(E-\frac{k^2}{2\mu}-\frac{p^2}{2\mu}\right)-i\eps \right)\:, \\
	&V^S_{4}(k,q)=-\frac{M_{D^*}}{k p}Q_0\left(-\frac{M_{D^*}}{k p}\left(E-\frac{k^2}{2\mu_*}-\frac{p^2}{2\mu}\right)-i\eps \right)\:. \notag
\end{align}

Note that the projections onto isospin $1/2$ are the same as that in the previous subsection, that is, $C_1^I=C_2^I=C_3^I=C_0^I=1$, and the additional spin projector for the $T^*_{cc} D$ scattering is given by
\begin{align}\label{eq: pw_operators}
	{\cal O}_{j, i}^{J=1}(1\otimes 0\to1)=\:&\delta_{ij}\:.
\end{align}
Using the same strategy as presented in the $T^*_{cc}D^*$ subsection, one can obtain the projection coefficients for
the $S$-wave $T^*_{cc}D$-$T_{cc}D^*$ coupled-channel integral equations by applying the above operators to Eqs.~\eqref{eq: TstarD-1} and \eqref{eq: TstarD-2}.
The results are collected in Table~\ref{Tab: coeff-TstarD}.
\begin{table}[tb]
	\centering
	\renewcommand\arraystretch{1.5}
	\caption{Coefficients of the partial-wave projected integral equation for the $S$-wave $T^*_{cc}D$ scattering.\label{Tab: coeff-TstarD}}
	\begin{tabular}{l  c  c  c}
		\hline
		\hline
		Channel & $C_1^IC_1^J$ & $C_2^IC_2^J$ & $C_3^IC_3^J$ \\
		\hline
		$(I,J)=(\frac12,1)$  & $1\times\sqrt{2}$  & $1\times \sqrt{2}$	&	$1\times(-1)$ \\
		\hline
		\hline
	\end{tabular}
\end{table}

\begin{figure}[tb]
    \centering
    \includegraphics[width=0.7\linewidth]{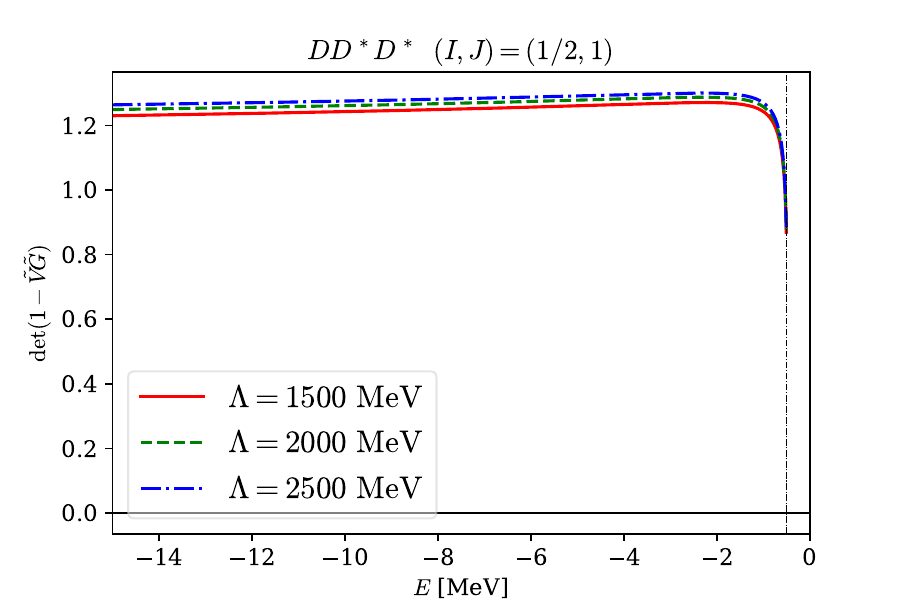}
    \label{fig:detD2channel}
    \caption{Determinant $\operatorname{det}(1-\tilde{V}\tilde{G})$ of the $T_{cc}^*D$-$T_{cc}D^*$ coupled-channel scattering equations with quantum numbers $(I,J)=(1/2,1)$. The vertical line denotes the threshold of $T_{cc}^*D$. }\label{fig:detD2channel}
\end{figure}

The coupled-channel integral equations are then expressed in the following general form with these spin-isospin factors as,
\begin{align}\label{eq: TstarD_pw}
	(T_1)_{(0)}^{I, J}(E,k,p) =&\int_0^{\Lambda} dq\: C_1^IC_1^J {\cal M}_{12}(p,q) (T_2)_{(0)}^{I, J}(E,k,q)\:,\notag\\
	(T_2)_{(0)}^{I, J}(E,k,p) =&\, C_2^IC_2^J\:{\cal M}_{20}(k,p) +\int_0^{\Lambda} dq\: C_2^IC_2^J {\cal M}_{21}(p,q) (T_1)_{(0)}^{I, J}(E,k,q)\notag\\
	&\qquad\qquad\qquad\quad\: +\int_0^{\Lambda} dq\: C_3^IC_3^J {\cal M}_{22}(p,q) (T_2)_{(0)}^{I, J}(E,k,q)\:.
\end{align}
The four scalar amplitudes ${\cal M}_{12}$, ${\cal M}_{20}$, ${\cal M}_{21}$, and ${\cal M}_{22}$ have been defined by
Eqs.~\eqref{eq: TstarD-1} and \eqref{eq: TstarD-2}.
The cutoff dependence of the determinant is shown in Fig.~\ref{fig:detD2channel}, and again no Efimov effect is observed. The behavior of the phase shifts and scattering lengths is similar to $D^*T^*_{cc}$ system in the previous section, and therefore will not be plotted here and below.

\subsection{$T_{cc} D^*$ scattering}

Finally, the amplitude for the $J=0$ or $J=2$ $S$-wave $T_{cc} D^*$ scattering satisfies the single-channel integral equation, as described above, which is given by 
\begin{align}\label{eq: TDstar}
	T_{(0) \, j i\alpha}^{\quad l k \beta} (E,k,p) =& -\frac{\pi \gamma_c}{\mu_c^2}\frac{2S_{3,3}}{S_{2,c}c_c} V^S_3(k,p)\:\boxed{\delta_{i l}\delta_{k j}\delta_{\beta\alpha}}\notag\\
	& - \frac{1}{\pi\mu_c}\frac{S_{3,3}}{S_{2,c}c_c} \int_0^{\Lambda} dq \frac{q^2 \: V^S_3(q,p) \: T_{(0) \, j i \alpha}^{\quad s n \mu} (E,k,q)\:\boxed{\delta_{n l}\delta_{k s}\delta_{\beta\mu}}}{-\gamma_c + \sqrt{-2\mu_c \left(E - \frac{q^2}{2M_{D^*}} - \frac{q^2}{2M_c} \right)
	- i\varepsilon}} \notag\\
	\equiv&\, {\cal M}_0\:\boxed{C_{4\;j i \alpha}^{\,\,\, lk \beta}}+\int_0^{\Lambda} dq\: {\cal M}_1 T_{(0) \, j i \alpha}^{\quad sn \mu} (E,k,q)\:\boxed{C_{4\; sn \mu}^{\,\,\,lk \beta}}.
\end{align}
After applying the partial wave projections, one obtains the prefactors as listed in Table~\ref{Tab: coeff-TDstar}.
\begin{table}[tb]
	\centering
	\renewcommand\arraystretch{1.5}
	\caption{Coefficients of the partial-wave projected integral equation for the $S$-wave $T_{cc}D^*$ scattering.\label{Tab: coeff-TDstar}}
	\begin{tabular}{l c c}
		\hline
		\hline
		Channel & $(I,J)=(1/2,0)$  & $(I,J)=(1/2,2)$ \\
		\hline
		$C_4^IC_4^J$  & $1\times1$ 	&	$1\times1$ \\
		\hline
		\hline
	\end{tabular}
\end{table}

\begin{figure}[tb]
    \centering
    \includegraphics[width=0.7\linewidth]{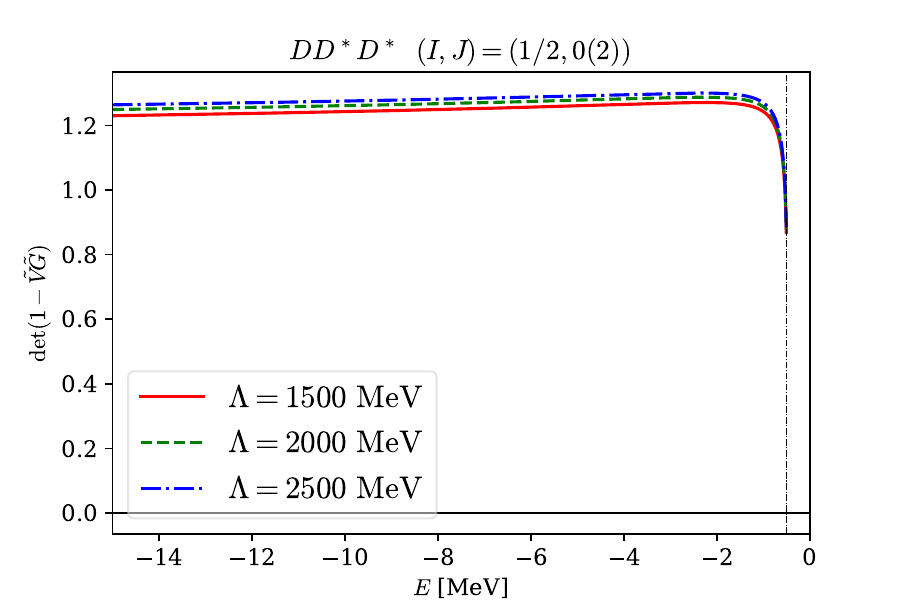}
    \label{fig:detD2s}
    \caption{Determinant $\operatorname{det}(1-\tilde{V}\tilde{G})$ of the $T_{cc}D^*$ single-channel scattering equations with quantum numbers $(I,J)=(1/2,0(2))$.
	The vertical line denotes the threshold of $T_{cc}D^*$.
	}
	\label{fig:detD2s} 
\end{figure}

Then the projected integral equation for the $S$-wave $T_{cc} D^*$ scattering reads
\begin{align}\label{eq: TDstar_pw}
	T_{(0)}^{I, J}(E,k,p) = C_4^IC_4^J\:{\cal M}_{0}(k,p)+\int_0^{\Lambda} dq\: C_4^IC_4^J {\cal M}_{1}(p,q) T_{(0)}^{I, J}(E,k,q)\:,
\end{align}
with two scalar amplitudes ${\cal M}_{0}$ and ${\cal M}_{1}$ defined by Eq.~\eqref{eq: TDstar}.
Once again, there is no signal of the Efimov effect in these channels, see Fig.~\ref{fig:detD2s}.

\section{$S$-wave $D^{*}D^*D^*$ system with dimers of all possible quantum numbers\label{sec:dis}}

In the preceding section, our analysis demonstrated that the Efimov effect in the coupled-channel Lippmann-Schwinger equation for three-charmed systems involving the $T^*_{cc}$ and $T_{cc}$ dimers is significantly affected by spin, isospin, and symmetry factors arising from identical particles. 
The weak cutoff dependence in the solutions of the scattering equations shows that there is no Efimov effect when only the $(I,J)=(0,1)$ dimers $T_{cc}$ and $T_{cc}^*$ are present. 
We will extend our analysis in this section to consider all possible dimer configurations and their implications for Efimov physics.
We will focus on the $D^*D^*D^*$ three-body system studied in Ref.~\cite{Ortega:2024ecy} to investigate under what conditions the Efimov effect can occur.

The Efimov effect emerges from the delicate interplay between short-range two-body interactions and long-range three-body correlations. It depends on the behavior at large momenta within the scattering equation, as exemplified by the three-body scattering equation for identical scalar isoscalar bosons~\cite{Bedaque:1998kg,Bedaque:1998km}. When the momentum $p$ is much larger than the inverse of the two-body scattering length, the scattering amplitude $T(p)$ satisfies
\begin{equation}\label{eq:TAs}
    T(p)=\frac{4}{\sqrt{3}\pi p}\int^{\infty}_{0}{dq}\:T(q)\ln{\frac{p^2+pq+q^2}{p^2-pq+q^2}}.
\end{equation}
By introducing the power-law ansatz $T(p)\sim p^{s-1}$, one obtains in the asymptotic momentum limit,
\begin{equation}
    p^s=\frac{4}{\sqrt{3}\pi}\int^{\infty}_{0}{dq}\:q^{s-1}\ln{\frac{p^2+pq+q^2}{p^2-pq+q^2}}.
\end{equation}
Consequently, a transcendental equation for the scaling parameter $s$ is obtained,
\begin{equation}\label{eq:Efimov condition}
    1=\frac{8}{\sqrt{3}}\frac{1}{s}\frac{\sin{\frac{\pi}{6}s}}{\cos{\frac{\pi}{2}s}}
\end{equation}
This equation has imaginary solutions $s_{0}=\pm 1.00624i$, indicating an oscillatory asymptotic behavior for $T(p)$. It becomes necessary to introduce a logarithmic periodic three-body contact interaction to achieve a unique solution as $\Lambda\to\infty$~\cite{Bedaque:1998kg,Bedaque:1998km}. 

This oscillatory behavior manifests as the Efimov effect, characterized by an infinite geometric spectrum of three-body bound states. When two particles interact with a resonant two-body interaction (infinite scattering length), a third particle can induce an effective attractive interaction, leading to the formation of weakly bound three-body states. These Efimov states follow a geometric scaling pattern, with each consecutive energy level related by a universal scaling factor of approximately $e^{2\pi/|s_0|} \approx 515$, as given in Eq.~\eqref{eq:Efimov scaling}. This remarkable phenomenon is universal, appearing in nuclear physics, atomic physics, and other quantum systems regardless of the underlying short-range interactions (for a review, see Ref.~\cite{Braaten:2004rn}).

When the bosons have spin and isospin, there will be various channels with different CG coefficients and symmetry factors. 
Then the asymptotic behavior of scattering equation will depend on these CG coefficients and symmetry factors. For coupled-channel scattering equations, if all scattering lengths are unnaturally large, Eq.~\eqref{eq:TAs} is generalized to 
\begin{equation}
T_i(p)=\mathcal{A}_{ij}\frac{4}{\sqrt{3}\pi p}\int^{\infty}_{0}{dq}\:T_j(q)\frac{1}{2}\ln{\frac{p^2+pq+q^2}{p^2-pq+q^2}} .
\label{eq:TAs_cc}
\end{equation}
The factor $\mathcal{A}_{ij}$ contributes to the coefficient of the scattering equations. One can decouple the above coupled-channel scattering equation into single-channel equations of ${\hat{T}}_i(p)$ by a unitary transformation $\cal S$, which diagonalizes $\cal A$ to $\cal {S}^{\mathrm{-1}}\mathcal{A} 
S=\rm diag(\lambda_{1},\lambda_{2},..)$, as (cf. Ref.~\cite{Hildenbrand:2019sgp})
\begin{align}
    \begin{pmatrix}
       T_{1}\\
       T_{2}\\
       ...
    \end{pmatrix}=  \cal S  \begin{pmatrix}
       \hat{T}_{1}\\
       \hat{T}_{2}\\
       ...
    \end{pmatrix}.
\end{align}
The behavior of each ${\hat{T}}_i(p)$ is governed by the corresponding eigenvalue $\lambda_i$ of $\mathcal{A}$. Using the ansatz ${\hat{T}}_i(p)\sim p^{s_i-1}$, we obtain equations for $s_i$:
\begin{equation}
    1=\frac{4\lambda_i}{\sqrt{3}}\frac{1}{s_i}\frac{\sin{\frac{\pi}{6}s_i}}{\cos{\frac{\pi}{2}s_i}}
\end{equation}
The Efimov effect occurs in the system when this equation has an imaginary solution, which happens when $\lambda_i>\lambda_c={3\sqrt{3}}/{(2\pi)}\approx 0.826993$~\cite{Braaten:2004rn}.
It is worth noting that for a system of three identical scalar bosons without (iso)spin, $\lambda=2$ as shown in Eq.~\eqref{eq:Efimov condition}.

It should be noted that the Bose-Einstein statistics plays a significant role in this context. The matrix $\mathcal{A}$ contains the symmetry factor $S_3$ from the one-boson exchange diagram, as defined in Eq.~\eqref{eq: tree_amp}, as well as the spin and isospin CG coefficients, the two-body symmetry factor $S_2$, and the normalization constant $c$ in $G_A$. We summarize the matrix $\mathcal{A}$ and its eigenvalues $\lambda_i$ for the $D^*D^*D^*$ system in Table~\ref{tab:A}.

\begin{table}[tb]
\caption{The $\mathcal{A}$ matrix and its eigenvalues $\lambda_i$ for the $D^*D^*D^*$ three-body scattering equation with various $(I,J)$ quantum numbers. The row and column indices of $\mathcal{A}_{ij}$ represent the three possible dimers with quantum numbers $(I,J)=(1,0)$, $(0,1)$, and $(1,2)$, respectively, when they contribute to the channel. 
Specifically, for $(I,J)=(1/2,1)$, all three dimers contribute to the coupled-channel system, so index $1$ corresponds to $(1,0)$, index $2$ to $(0,1)$, and index $3$ to $(1,2)$. 
For $(I,J)=(3/2,1)$, only dimers with $(I,J)=(1,0)$ and $(1,2)$ contribute, so we use index $1$ for $(1,0)$ and index $2$ for $(1,2)$.
For $(I,J)=(1/2,2)$, only dimers with $(I,J)=(0,1)$ and $(1,2)$ contribute, so we use index $1$ for $(0,1)$ and index $2$ for $(1,2)$. The case of $(I,J)=(3/2,0)$ is forbidden by the Bose-Einstein statistics.}
\begin{ruledtabular}
\begin{tabular}{l c c}
$(I,J)$&$\mathcal{A}$&$\lambda_{i}$\\
\hline
$(1/2,0)$ & $-1$&$-1$\\
\hline
$(3/2,0)$ & -&-\\
\hline
$(1/2,1)$ & $\begin{pmatrix}
    -\frac{1}{3}&1&-\frac{\sqrt{5}}{3}\\
    1&\frac{1}{2}&-\frac{\sqrt{5}}{2}\\
    -\frac{\sqrt{5}}{3}&-\frac{\sqrt{5}}{2}&-\frac{1}{6}
\end{pmatrix}$&$2,-1,-1$\\
\hline
$(3/2,1)$ & $\begin{pmatrix}
    \frac{2}{3}&\frac{2\sqrt{5}}{3}\\
    \frac{2\sqrt{5}}{3}&\frac{1}{3}
\end{pmatrix}$&$2,-1$\\
\hline
$(1/2,2)$ & $\begin{pmatrix}
    \frac{1}{2}&\frac{3}{2}\\
    \frac{3}{2}&\frac{1}{2}
\end{pmatrix}$&$2,-1$\\
\hline
$(3/2,2)$ & $-1$& $-1$\\
\hline
$(1/2,3)$ & $-1$& $-1$\\
\hline
$(3/2,3)$ & $2$& $2$
\end{tabular}
\end{ruledtabular}
\label{tab:A}
\end{table}

Therefore, for the $D^*D^*D^*$ system, we conclude in Sec.~\ref{sec:NREFT} that no Efimov effect occurs when only the existence of the $T^*_{cc}$ dimer with quantum numbers $(I,J)=(0,1)$ is assumed.
However, when dimers with quantum numbers $(I,J)=(1,2)$ and $(1,0)$ are included in Eq.~\eqref{eq:TAs_cc} (that is, these two-body $D^*D^*$ channels approach the unitary limit), the Efimov effect emerges in channels where the eigenvalue equals 2, exceeding the critical threshold $\lambda_{c}$. In these cases, the scaling factor $s_{i}$ for the Efimov states equals $\pm1.00624i$, which precisely matches the value observed in systems of three identical spinless and isoscalar bosons.
In particular, for the $(I,J)=(1/2,2)$ system, the Efimov effect can emerge if, in addition to the existence of the isoscalar $T^*_{cc}$, the $D^*D^*$ scattering length in the $(I,J)=(1,2)$ channel is sufficiently large.

Let us take the case of $(I,J)=(1/2,1)$ in Table \ref{tab:A} as an example. 
If we set all three $D^*D^*$ dimers $(1,0)$, $(0,1)$, and $(1,2)$ with binding energies equal to $0.50$ MeV (corresponding to the binding energy of $T^*_{cc}$), when $\Lambda=1500$ MeV we can find the first two bound states located at $E_1=-44.3$~MeV and $E_2=-0.83$~MeV with a vanishing three-body force. It is noteworthy that $\Delta E_1/\Delta E_2\approx 134$ with $\Delta E_{n}=E_{n}+E^{*}_{B}$, which is similar to the result with the same value of binding energy in Ref.~\cite{Ortega:2024ecy}.

\section{Summary and Discussions}

The existence of the $T_{cc}(3875)^+$ extremely close to the $DD^*$ threshold implies a large $S$-wave scattering length in the isoscalar $DD^*$ channel. Motivated by this observation and the hypothetical heavy quark spin partner of the $T_{cc}$, denoted as $T_{cc}^*$ with quantum numbers $(I,J)=(0,1)$, we have investigated the possibility of the Efimov effect in the $D^*D^*D^*$ and $DD^*D^*$ three-body systems.
Our analysis reveals that if only the isoscalar $J=1$ dimers $T_{cc}$ and $T_{cc}^*$ exist, no Efimov effect is expected. This conclusion differs from that of Ref.~\cite{Ortega:2024ecy}, which reported the existence of $(I,J)=(1/2,0)$ $D^*D^*D^*$ three-body bound states due to the inconsistent prefactors arising from the partial wave projection. 

Then we investigate under what conditions there can be Efimov effects in the $S$-wave $D^*D^*D^*$ system. 
We find that if the unitary limit is approached also in the isovector $(I,J)=(1,0)$ and $(I,J)=(1,2)$ $D^*D^*$ channels, the Efimov effect can be reinstated. 
Notice that such a case corresponds to one of the solutions of minimizing the entanglement~\cite{Beane:2018oxh} for the $S$-wave $D^{(*)}D^{(*)}$ scatterings~\cite{Hu:2024hex}. However, since the partial decay width of the $T_{cc}^*\to DD^*$ channel could reach around 20~MeV~\cite{Dai:2021vgf}, Efimov states in $D^*D^*D^*$---if they exist at all by satisfying the conditions mentioned above---will be hard to be found in experiments.
It would be interesting to calculate the isovector $D^*D^*$ low-energy scatterings with $J=1$ and 2 using lattice QCD.

\begin{acknowledgments}
    This work is supported in part by the National Key R\&D Program of China under Grant No. 2023YFA1606703; by the National Natural Science Foundation of China (NSFC) under Grants No. 12125507, No. 12361141819,  and No. 12447101; by the Chinese Academy of Sciences (CAS) under Grant No. YSBR-101. U.-G.M. and A.R. are also partially supported by the CAS  President's International Fellowship Initiative (PIFI) under Grant Nos. 2025PD0022 and 2024VMB0001, respectively, and AR in addition acknowledges the financial support
from the Ministry of Culture and Science of North Rhine-Westphalia through the
NRW-FAIR program. H.-L.F. thanks the University of Bonn and Y.-H.L. thanks the Institute of Theoretical Physics, CAS,  for the hospitality during their visits.
    H.-W.H. was supported by Deutsche Forschungsgemeinschaft (DFG, German Research Foundation) under Project ID 279384907 – SFB 1245 and by the German Federal Ministry of Education and Research (BMBF) (Grant No. 05P24RDB).
\end{acknowledgments}

\bibliography{refs.bib}

\end{document}